\documentclass[conference]{IEEEtran}
\usepackage{amsfonts}
\usepackage{amsmath, amssymb}
\usepackage{graphics,graphicx}
\usepackage{epsfig}
\usepackage{color}
\newtheorem{thm}{Theorem}

\newtheorem{lem}{Lemma}

\newtheorem{rem}{Remark}
\IEEEoverridecommandlockouts

\renewcommand{\baselinestretch}{1} % Uncomment for 1.5 spacing between lines
\begin{document}
\title{A Layered Lattice Coding Scheme for a Class of Three User Gaussian Interference Channels}

\author{Sriram Sridharan, Amin Jafarian, Sriram Vishwanath, Syed A. Jafar and Shlomo Shamai (Shitz)
\thanks{S. Sridharan, A. Jafarian and S. Vishwanath are with the Wireless Networking and Communications Group, Department of Electrical and Computer Engineering, University of Texas at Austin, Austin, TX - 78712 (email: sridhara@ece.utexas.edu; jafarian@ece.utexas.edu; sriram@ece.utexas.edu). S. Sridharan, A. Jafarian and S. Vishwanath are supported in part by National Science Foundation
grants NSF CCF-0448181, NSF CCF-0552741, NSF CNS-0615061, and NSF
CNS-0626903, THECB ARP and the Army Research Office YIP.}%
\thanks{S. A. Jafar is with the Department of Electrical and Computer Science, University of California at Irvine, Irvine, CA 96297 (email : syed@ece.uci.edu). the work of S. A. Jafar was supported by ONR Young Investigator Award N00014-08-1-0872}%
\thanks{S. Shamai (Shitz) is with the Department of Electrical Engineering, Technion-Israel Institute of Technology, Technion City, Haifa 32000, Israel (email : sshlomo@ee.technion.ac.il). The work of S. Shamai was supported
by the Israel Science Foundation.}%
}
\maketitle
\begin{abstract}
 The paper studies a class of three user Gaussian interference channels. A new layered lattice coding scheme is introduced as a transmission strategy. The use of lattice codes allows for an ``alignment" of the interference observed at each receiver. The layered lattice coding is shown to achieve more than one degree of freedom for a class of interference channels and also achieves rates which are better than the rates obtained using the Han-Kobayashi coding scheme. 
  %In this extended abstract, a new achievable region is determined for a class of $K >2$ user Gaussian interference channels. The achievable scheme brings together lattice codes with the more traditional Han-Kobayashi  transmission techniques for this channel. The use of lattice codes instead of random codes allows for an ``alignment" of the interference observed by each receiver. We demonstrate numerically that this alignment results in an improvement in the sum rate achievable over that achieved using the Han-Kobayashi strategy.
\end{abstract}

\section{Introduction}
Determining the capacity region of the Gaussian interference channel (IC) has been a long standing open problem. Several achievable regions and outer bounds have been obtained over the last three decades \cite{Carleial1978}--\nocite{Benzel1979, HanKobayashi1981, Sato1981, GamalCosta1982, Costa1985, EtkinTseWang2007, ShangKramerChen2007, AnnapureddyVeeravalli}\cite{MotahariKhandani}. Recently, it has been shown in \cite{EtkinTseWang2007} that the Han-Kobayashi achievable region \cite{HanKobayashi1981} achieves the capacity region of the two user Gaussian IC to within one bit. However, this result does not extend to an IC with more than two users. In \cite{ShangKramerChen2007}--\nocite{AnnapureddyVeeravalli}\cite{MotahariKhandani}, the sum capacity of the two user Gaussian IC has been determined for a range of ``weak" interference cases.

For interference networks with more than two transmitter-receiver pairs, degrees of freedom characterization (capacity approximations within $o(\log(\mbox{SNR}))$) have been found for a class of time or frequency varying channels in \cite{CadambeJafarInt}--\nocite{CadambeJafarWirelessXNetworks, CadambeJafarWirelessNetworks}\cite{ShenMadsenVidal}. These results do not apply to interference networks with channels that are not time or frequency varying. For fixed interference networks, very little is known even about its degrees of freedom characterization. In \cite{CadambeJafarShamai}, some examples of $K$ user ICs are presented which come close to achieving $K/2$ degrees of freedom. However, for general $K$ user ICs with fixed channel gains, only one degree of freedom has been achieved, whereas the best outer bound on the total degrees of freedom is $K/2$ as derived in \cite{MadsenNosratinia}. 
The fundamental problem, we believe, is the achievable strategy that forms the baseline of study - the Han-Kobayashi scheme \cite{HanKobayashi1981}. In this paper, we use a layered lattice coding scheme to provide rate improvements in a three user IC over what can be obtained using the layered random coding scheme used in \cite{HanKobayashi1981}. We also show that the layered lattice coding scheme can achieve more than one degree of freedom for a wide class of three user ICs. The results of the paper can be extended to $K$ user ICs.

The use of structured codes for ICs has already been introduced for one-sided ICs in \cite{BreslerParekhTse} and more recently in \cite{Sridharan-Globecom} in determining capacity of very-strong ICs. Lattice coding has also been used as an effective transmission strategy in achieving the capacity of several other channels. It is used as a transmission strategy for an AWGN channel in \cite{deBuda}--\nocite{Poltyrev}\cite{Loeliger} to achieve a rate close to capacity in the high SNR regime. In \cite{ErezZamir2004}, lattice coding is shown to achieve the capacity of the AWGN channel. Some other relevant results include \cite{ErezLitsynZamir}--\nocite{ZamirShamaiErez, ErezShamaiZamir, NamElGamal, KrithivasanPradhan, NazerGastparAllerton2007}\cite{NazerGastpar}.

The rest of the paper is organized as follows: In Section II, we describe the system model for the three user Gaussian ICs and introduce the class of channels we analyze in this paper. In Section III, we introduce preliminaries on lattice coding and summarize some relevant results that we will use in this paper. In Section IV, we present the layered lattice coding approach for a symmetric three user Gaussian IC from a degree of freedom perspective. In Section V, we describe an achievable region for the symmetric three user Gaussian IC. In Section VI, we generalize the results of Sections IV to  a class of non-symmetric channels. We conclude in Section VII.

\section{System Model}
We consider the three user Gaussian IC with three transmitter-receiver pairs and three independent messages, where message $m_k$ originates at transmitter $k$ and is intended for receiver $k$, for $k \in \{1,2,3\}$. The system model is shown in Figure $1$ and the channel equations are described by
\begin{equation}
Y_j(i) = X_j(i) + \sum_{k=1, k \neq j}^3 h_{jk} X_k(i) + Z_j(i), \ \forall \ j \in \{1,2,3\},
\end{equation}
\begin{figure}[hbtp]
\centering
\includegraphics[width=3 in]{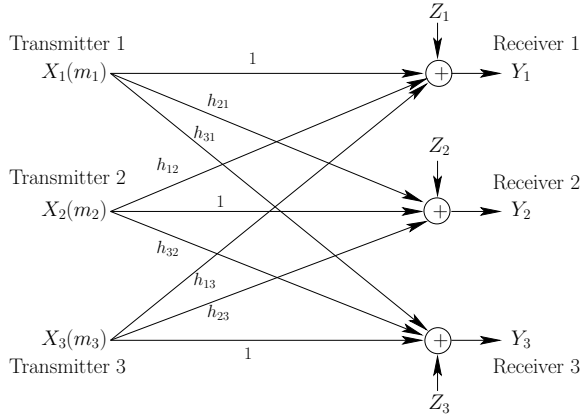}
\caption{System Model for Three User Gaussian Interference Channel}
\label{fig : SystemModel}
\end{figure}
where $Y_j(i)$ is the received signal at the $j^{th}$ receiver at the $i^{th}$ channel use, $X_k(i)$ is the transmitted signal at the $k^{th}$ transmitter at the $i^{th}$ channel use, and $h_{jk}$ denotes the channel gain from the $k^{th}$ transmitter to the $j^{th}$ receiver. In the above equation, all the direct channel gains have been normalized to unity. $Z_j(i)$ is the zero mean, unit variance additive white Gaussian noise at receiver $j$ at time $i$. The Gaussian noise at each receiver is i.i.d. across time.
The channel inputs are subject to the following average power constraints:
\begin{equation}
\frac{1}{n}\sum_{i=1}^n X_k(i)^2 \leq P_k, \ \forall\ k \in \{1,2,3\}.
\end{equation}
Let $H$ denote the $3 \times 3$ matrix of channel gains
\begin{displaymath}
H = \left(\begin{array}{ccc}1 & h_{12} & h_{13} \\ h_{21} & 1 & h_{23}\\ h_{31} & h_{32} & 1\end{array}\right).
\end{displaymath}
Let $\mathcal{H}_1$ denote a class of channel matrices such that
\begin{equation}\label{eqn : channel class}
\mathcal{H}_1 = \left\{\begin{array}{c} H \in \mathbb{R}^{3\times 3} : h_{11} = h_{22} = h_{33} = 1,\\
 \textrm{ and } \frac{h_{12}}{h_{21}}\times\frac{h_{23}}{h_{32}}\times\frac{h_{31}}{h_{13}} \in \mathbb{Q} \end{array}\right\},
\end{equation}
In this work, we will restrict ourselves to three-user Gaussian ICs whose channel matrices are in the set $\mathcal{H}_1$. This is the class of channels to which our coding strategy applies. Note that this is a (highly) non-trivial class of channels which includes the symmetric IC as a special case. Achievable rates and capacity region are defined in the Shannon sense.
The total degrees of freedom of the three user Gaussian IC is denoted by $\mathcal{D}_{sum}$ and is defined as
\begin{equation}\label{eqn : total degrees of freedom definition}
\mathcal{D}_{sum} \triangleq \limsup_{P_1 + P_2 + P_3 \rightarrow \infty}\max_{(R_1, R_2, R_3)\in \mathcal{C}_{ap}} \frac{R_1 + R_2 + R_3}{\frac{1}{2}\log(P_1+P_2+P_3)}.
\end{equation}
The total degrees of freedom represents the rate of growth of sum capacity in terms of $\log(\textrm{SNR})$.

\section{Lattice Coding Preliminaries}
In this section, we review some preliminaries on lattice coding and summarize some results on lattice coding that we use in this paper. A lattice $\Lambda$ of dimension $n$ is described by
\begin{displaymath}
\Lambda = \{\lambda = Gx : x \in \mathbb{Z}^n\},
\end{displaymath}
where $G$ is the generator matrix that describes the lattice. Let $\Omega_{\Lambda}$ denote the fundamental Voronoi region of the lattice $\Lambda$ and $V_{\Lambda}$ denote the volume of $\Omega_{\Lambda}$. In this paper, we use lattices generated using construction A described below (used in \cite{Loeliger}).

For any positive integer $p$, let $\mathbb{Z}_p$ denote the set of integers modulo $p$. Let $g : \mathbb{Z}^n \rightarrow \mathbb{Z}_p^n$ denote the componentwise modulo $p$ operation over integer vectors. Let $C$ denote a linear $(n,k)$ code over $\mathbb{Z}_p$. Then the lattice $\Lambda_C$ given by
\begin{equation}\label{eqn : construction A}
\Lambda_C = \left\{v \in \mathbb{Z}^n : g(v) \in C\right\}
\end{equation}
is generated using construction A with respect to the linear code $C$. 
A set $\mathcal{B}$ of linear codes over $\mathbb{Z}_p$ is ``balanced" if every nonzero element of $\mathbb{Z}_p^n$ is contained in the same number of codes in $\mathcal{B}$. Let $\mathcal{L}_B$ be the set of lattices denoted by
\begin{equation}\label{eqn : lattice set}
\mathcal{L}_{B} = \{\Lambda_C : C \in \mathcal{B}\}.
\end{equation}
We consider a single user point to point additive noise channel
\begin{equation}\label{eqn : Gaussian channel}
Y = X + Z
\end{equation}
where $X$ is the transmitted signal, $Y$ the received signal and $Z$ is the additive noise of zero mean and variance equal to $\sigma^2$ that corrupts the transmitted signal at the receiver. If the transmitted word over time is a lattice point, then it can be shown that a suitable lattice and a decoding strategy exists such that the probability of decoding error can be made arbitrarily small as the number of dimensions of the lattice increases. This result is stated formally in the following lemma.
\begin{lem}[\cite{Loeliger}]\label{lem : Loeliger}
Consider a single user point to point additive noise channel described in (\ref{eqn : Gaussian channel}). Let $\mathcal{B}$ be a balanced set of linear $(n,k)$ codes over $\mathbb{Z}_p$. Averaged over all lattices from the set $\mathcal{L}_{B}$ given in (\ref{eqn : lattice set}), each scaled by $\gamma > 0$ and with a fundamental volume $V$, we have that for any $\delta > 0$, the average probability of decoding error is bounded by
\begin{equation}
\overline{P_e} < (1 + \delta) \frac{2^{n \frac{1}{2}\log(2\pi e \sigma^2)}}{V}.
\end{equation}
for sufficiently large $p$ and small $\gamma$ such that $\gamma^n p^{n-k} = V$. Hence, the probability of decoding error for at least three fourths of the lattices in $\mathcal{L}_{B}$ satisfies
\begin{equation}\label{eqn : Loeliger condition}
Pe < 4(1 + \delta) \frac{2^{n \frac{1}{2}\log(2\pi e \sigma^2)}}{V}.
\end{equation}
\end{lem}
The proof of the lemma is described in \cite{Loeliger} and is omitted here. The next lemma describes the existence of a good lattice code for a single user AWGN channel.
\begin{lem}
Consider a single user point to point additive noise channel described in (\ref{eqn : Gaussian channel}) such that the transmitter satisfies a power constraint of $P$. Then, we can choose a lattice $\Lambda$ generated using construction A, a shift $s$ and a shaping region $S$ such that the codebook $(\Lambda + s) \cap S$ achieves a rate $R$ with arbitrarily small probability of error if
\begin{displaymath}
R \leq \frac{1}{2}\log\left(\frac{P}{\sigma^2}\right).
\end{displaymath}
\end{lem}
The proof of the lemma is described in \cite{Loeliger}. In essence, the above lemma describes the existence  of a lattice code with sufficient codewords. It should be noted that lattice coding with a random dither has been used in \cite{ErezZamir2004} to achieve the full capacity of the AWGN channel. In the next section, we describe a layered lattice coding approach for a symmetric three user Gaussian IC from degree of freedom perspective.

\section{Symmetric IC: Degrees of Freedom}
In this section, we analyze the degrees of freedom of three user symmetric Gaussian IC. For this channel, the channel gain matrix is of the form
\begin{displaymath}
H = \left(\begin{array}{ccc}1 & a & a\\ a & 1 & a\\ a & a & 1\end{array}\right).
\end{displaymath}
We analyze the degrees of freedom of such a channel using a layered lattice coding approach. First, we briefly review the ``very strong" interference result of \cite{Sridharan-Globecom} for a symmetric three user Gaussian IC in the next lemma.

\begin{lem}\cite[Theorem 2]{Sridharan-Globecom}\label{lem : very strong interference}
Consider a symmetric three user Gaussian IC with channel gain $a$, power constraint $P$ at the transmitters and noise variance $\sigma^2$ at the receivers. If the channel gain satisfies $a^2 \geq \frac{P}{\sigma^2} + 1$, then each user can achieve a symmetric rate of $\frac{1}{2}\log\left(\frac{P}{\sigma^2}\right)$.
\end{lem}
The proof of Lemma \ref{lem : very strong interference} for $\sigma^2 = 1$ is given in \cite[Theorem 2]{Sridharan-Globecom}. The proof for any $\sigma^2$ follows with little modifications. We use this lemma frequently in the next Theorem which shows that the total degrees of freedom of a symmetric three user Gaussian IC is more than $1$ for a wide range of channel parameter $a$.

\begin{figure}[hbtp]
\label{fig : dof symmetric}
\centering
\includegraphics[width=3.5 in]{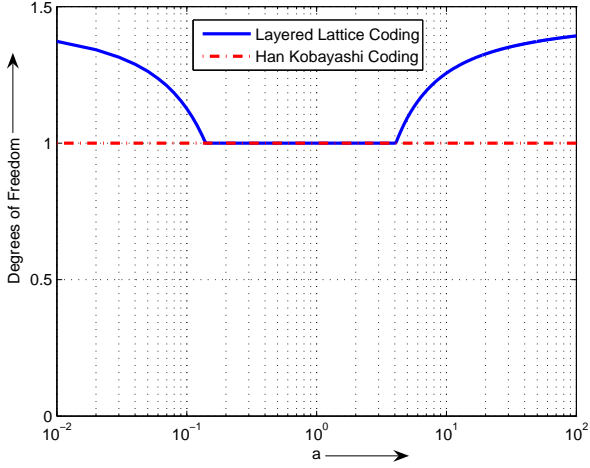}
\caption{Plot of Achievable Degrees of Freedom versus $a^2$}
\end{figure}

\begin{thm}\label{thm : degree of freedom of symmetric channel}
Consider a symmetric three user Gaussian IC with channel parameter $a$, and noise variance equal to $1$. The total degrees of freedom of the channel satisfies
\begin{equation}\label{eqn : Theorem 1}
\mathcal{D}_{sum} \geq \left\{\begin{array}{ll} \max\left(1, 3\times\frac{\log(a^2-1)}{\log(2a^4-a^2)}\right), &\ a^2 \geq 2\\
1, & \frac{1}{3} \leq a^2 \leq 2\\
\max\left(1, 3\times\frac{\log\left(\frac{1-a^2}{2a^2}\right)}{\log\left(\frac{1+a^2}{2a^4}\right)}\right), &\ a^2 \leq \frac{1}{3}
\end{array}\right.
\end{equation}
\end{thm}
\begin{proof}\textit{: }
The proof of the Theorem for $\frac{1}{3} \leq a^2 \leq 2$ is obvious, because a simple time sharing scheme will achieve one degree of freedom for any $a$. Hence, we focus on the other two cases.

First, we consider the case $a^2 \geq 2$. As the channel is symmetric and we are analyzing the total degrees of freedom, we look at only symmetric rate points. For $j \in \{1,2,3\}$, transmitter $j$ communicates message $m_j \in \{1, \ldots, 2^{nR}\}$ to receiver $j$. Transmitter $j$ splits its message $m_j$ into $N$ parts $m_{j1}, \ldots, m_{jN}$, such that a rate $R_i$ is associated with the $i^{th}$ sub-message of each message. For $i \in \{1,\ldots,N\}$, the $i^{th}$ sub-message is encoded to codeword $X_{ji}^n$ by the $j^{th}$ transmitter, which transmits $X_j^n = \sum_{i=1}^N X_{ji}^n$. Also, each transmitter assigns a power $P_i$ for encoding its $i^{th}$ sub-message. Note that the subscript in rate and power does not indicate user, but the sub-messages. The power $P_i$ is chosen as
\begin{equation}\label{eqn : power assignment 1}
P_i = (a^2 - 1)(2a^4-a^2)^{N-i}, \ \ i\in\{1,2,\ldots,N\}.
\end{equation}
We explain the encoding and decoding strategy below in detail.

Encoding Strategy: Each transmitter encodes all its sub-messages using lattice coding, and chooses lattices $\Lambda_1, \ldots, \Lambda_N$, shifts $s_1, \ldots, s_N$ and spherical shaping regions $S_1, \ldots, S_N$. The codebook for $i^{th}$ sub-message at each transmitters is denoted by $\mathcal{C}_i = (\Lambda_i + s_1) \cap S_i$. 
%That is, the codeword transmitted by transmitter $j$ for the $i^{th}$ sub-message, $X_{ji}^n \in \mathcal{S}_i$.

Decoding Strategy: The received signal at receiver $j$ is 
\begin{displaymath}
Y_j^n = \sum_{i=1}^N X_{ji}^n + \sum_{k=1, k \neq j}^3\sum_{i=1}^N a X_{ki}^n + Z_j^n.
\end{displaymath}
We denote the interference at receiver $j$ due to the $i^{th}$ sub-message from the other transmitters by $I_{ji}^n$ given by
\begin{equation}
I_{ji}^n = \sum_{k=1, k \neq j}^3 a X_{ki}^n.
\end{equation}
The decoding process at receiver $j$ proceeds through $N$ stages. At stage $i$, receiver $j$ first decodes interference $I_{ji}^n$ and then decodes message $m_{ji}$. In decoding the interference $I_{ji}^n$, receiver $j$ sees interference plus noise of
\begin{displaymath}
\sum_{k=i}^{N} X_{jk}^n + \sum_{l=1,l\neq j}^{3}\sum_{k=i+1}^{N} a X_{lk}^n + Z_j^n
\end{displaymath}
with an interference plus noise power $\leq P_i + \sum_{k=i+1}^N (2a^2 + 1)P_k + 1$. In decoding message $m_{ji}$, receiver $j$ sees an interference plus noise
\begin{displaymath}
\sum_{k=i+1}^{N} X_{jk}^n + \sum_{l=1,l\neq j}^{3}\sum_{k=i+1}^{N} a X_{lk}^n + Z_j^n
\end{displaymath}
with an interference plus noise power $\leq \sum_{k=i+1}^N (2a^2 + 1)P_k + 1$. Next, we describe the choice of lattices, shifts and spherical regions, before proceeding to probability of error analysis and rate constraints at the receivers.

Choice of Lattices, Shifts and Shaping Regions: For $i \in \{1,\ldots, N\}$, each transmitter chooses shaping region $S_i$ to be an $n$ dimensional sphere of radius $\sqrt{nP_i}$. The volume of the shaping region $S_i$ is denoted by $V_{S_i}$. Lattice $\Lambda_i$ is generated using construction A such that
\begin{itemize}
%\item Condition (\ref{eqn : MH condition}) (Minkowski-Hlawka condition) is satisfied,
\item the volume of the Voronoi region $V_i = 2^{-nR_i}V_{S_i}$,
\item in decoding interference $I_{ji}^n$ at receiver $j$, the probability of error is upper bounded by (\ref{eqn : Loeliger condition}) with $\sigma^2 = P_i + \sum_{k=i+1}^N (2a^2 + 1)P_k + 1$, and
\item in decoding message $m_{ji}$ at receiver $j$, the probability of error is upper bounded by (\ref{eqn : Loeliger condition}) with $\sigma^2 = \sum_{k=i+1}^N (2a^2 + 1)P_k + 1$.
\end{itemize}
Finally, shift $s_i$ is chosen such that the cardinality of the codebook $\mathcal{C}_i$ satisfies $|\mathcal{C}_i| = |(\Lambda_i + s_i) \cap S_i| \geq 2^{nR_i}$. Next, we describe the probability of error analysis and rate constraints at receiver $1$. The analysis and the rate constraints at other receivers are the same.

Receiver $1$ first decodes interference $I_{11}^n$ and message $m_{11}$. The interference plus noise power when decoding $I_{11}^n$ is given by
$P_1 + \sum_{k=2}^N (2a^2 + 1)P_k + 1$. With the choice of lattice $\Lambda_1$, the probability of decoding error is upper bounded by
\begin{equation}
P_{e1}^{int} \leq 4(1 + \delta) \frac{2^{n \frac{1}{2}\log(2\pi e (P_1 + \sum_{k=2}^N (2a^2 + 1)P_k + 1))}}{a^nV_1},
\end{equation}
where $a^n V_1$ is the volume of the Voronoi region of the lattice $a \Lambda_1$ (the interference lattice of message $m_{21}$ and $m_{31}$). Hence, the probability of error decays with $n$ if
\begin{equation}
R_1 \leq \frac{1}{2}\log\left(\frac{a^2 P_1}{P_1 + \sum_{k=2}^N (2a^2 + 1)P_k + 1}\right).
\end{equation}
Similarly, in decoding the message $m_{11}$, the interference plus noise power seen by receiver $1$ is equal to $\sum_{k=2}^N (2a^2 + 1)P_k + 1$. The probability of decoding error is upper bounded by
\begin{equation}
P_{e1}^{message} \leq 4(1 + \delta) \frac{2^{n \frac{1}{2}\log(2\pi e (\sum_{k=2}^N (2a^2 + 1)P_k + 1))}}{V_1}.
\end{equation}
Hence, the probability of error decays with $n$ if
\begin{equation}
R_1 \leq \frac{1}{2}\log\left(\frac{P_1}{\sum_{k=2}^N (2a^2 + 1)P_k + 1}\right).
\end{equation}
Proceeding along similar lines, at stage $i$, interference $I_{1i}^n$ and message $m_{1i}$ can be decoded successfully if
\begin{eqnarray}
R_i \leq \frac{1}{2}\log\left(\frac{a^2 P_i}{P_i + \sum_{k=i+1}^N (2a^2 + 1)P_k + 1}\right),\\
R_i \leq \frac{1}{2}\log\left(\frac{P_i}{\sum_{k=i+1}^N (2a^2 + 1)P_k + 1}\right).
\end{eqnarray}
The power values have been chosen so that the ``very strong" interference condition is satisfied at each stage. The noise plus interference power seen at stage $i$ in decoding interference $I_{1i}^n$ and message $m_{1i}$ is equal to $\sum_{k=i+1}^N (2a^2 + 1)P_k + 1$. From the power assignments in (\ref{eqn : power assignment 1}), we can see that
\begin{displaymath}
a^2 = \frac{P_i}{\sum_{k=i+1}^N (2a^2 + 1)P_k + 1} + 1.
\end{displaymath}
With the choice of power values as in (\ref{eqn : power assignment 1}), the rate at each stage is given by
\begin{equation}
R_i = \frac{1}{2}\log(a^2 - 1).
\end{equation}
For $R_i$ to be positive, we need $a^2 \geq 2$. Hence, the total rate achieved by each user is given by
\begin{equation}
R = \frac{1}{2}\log(a^2-1)^N.
\end{equation}
Also, the total power used by each transmitter is given by
\begin{eqnarray}
P &=& P_1 + \ldots + P_N\nonumber\\
%   &=& (a^2-1)\frac{(2a^4-a^2)^N-1}{2a^4-a^2-1}\\
   &\leq & (2a^4-a^2)^N.
\end{eqnarray}
Taking $N$ to $\infty$, we get the desired result. That is,
\begin{equation}
\lim_{N, P \rightarrow \infty} \frac{3 R}{\frac{1}{2}\log(P)} \geq 3\times\frac{\log(a^2-1)}{\log(2a^4-a^2)}.
\end{equation}
Next, we consider the case $a^2 \leq \frac{1}{3}$. The proof for this case is very similar to that of $a^2 \geq 2$ with very few modifications. We again focus only on symmetric rates. For $j \in \{1,2,3\}$, transmitter $j$ splits its message $m \in \{1,\ldots,2^{nR}\}$ into $N$ sub-parts $m_{j1}, \ldots, m_{jN}$ such that rate $R_i$ and power $P_i$ is associated with the $i^{th}$ sub-message. Power $P_i$ is chosen as
\begin{equation}\label{eqn : Power assignment 2}
P_i = \frac{1-a^2}{2a^4}\left(\frac{1+a^2}{2a^4}\right)^{N-i}.
\end{equation}
The encoding strategy is similar to the one described for the case $a^2 \geq 2$ in that each transmitter uses lattice coding to encode all its sub-messages. However, the decoding strategy is slightly different. The decoding process again proceeds through $N$ stages. In stage $i$, receiver $j$ first decodes message $m_{ji}$ and then decodes interference $I_{ji}^n$. This is because decoding interference first will lead to rate constraints that are more binding than the constraints due to decoding the message.

Choice of Lattices, Shifts and Shaping Regions: For $i \in \{1,\ldots, N\}$, each transmitter chooses shaping region $S_i$ to be a $n$ dimensional sphere of radius $\sqrt{nP_i}$. Lattice $\Lambda_i$ is generated using construction A such that
\begin{itemize}
%\item Condition (\ref{eqn : MH condition}) (Minkowski-Hlawka condition) is satisfied,
\item the volume of the Voronoi region $V_i = 2^{-nR_i}V_{S_i}$,
\item in decoding interference $I_{ji}^n$ at receiver $j$, the probability of error is upper bounded by (\ref{eqn : Loeliger condition}) with $\sigma^2 = \sum_{k=i+1}^N (2a^2 + 1)P_k + 1$, and
\item in decoding message $m_{ji}$ at receiver $j$, the probability of error is upper bounded by (\ref{eqn : Loeliger condition}) with $\sigma^2 = 2a^2 P_i + \sum_{k=i+1}^N (2a^2 + 1)P_k + 1$.
\end{itemize}
Finally, shift $s_i$ is chosen such that the cardinality of the codebook $\mathcal{C}_i$ satisfies $|\mathcal{C}_i| = |(\Lambda_i + s_i) \cap S_i| \geq 2^{nR_i}$. The details of the probability of error analysis are similar to the case $a^2 \geq 2$ and are omitted here. Using the power assignments in (\ref{eqn : Power assignment 2}), we see that each user achieves a rate $R_i$ for its $i^{th}$ sub message given by
\begin{equation}
R_i = \frac{1}{2}\log\left(\frac{1-a^2}{2a^2}\right).
\end{equation}
Hence, each user achieves a total rate $R$ given by
\begin{equation}
R = \frac{1}{2}\log\left(\frac{1-a^2}{2a^2}\right)^N.
\end{equation}
The total power expended by each transmitter is given by
\begin{eqnarray}
P &=& P_1 + \ldots + P_N\nonumber\\
  %&=& \frac{1-a^2}{2a^4}\frac{\left(\frac{1+a^2}{2a^4}\right)^N - 1}{\left(\frac{1+a^2}{2a^4}\right)-1}\\
  &\leq & \left(\frac{1+a^2}{2a^4}\right)^N.
\end{eqnarray}
Taking $N$ to $\infty$, we get the desired result.
This completes the proof of Theorem 1.
\end{proof}
In Figure \ref{fig : dof symmetric}, we plot the degrees of freedom that we achieve for a symmetric three user Gaussian IC using the layered lattice coding approach.
\begin{rem}
From (\ref{eqn : Theorem 1}), we can see that the achievable total degrees of freedom tends to $3/2$ as the channel gain $a \rightarrow \infty$, and when $a \rightarrow 0$. We should also see that when the channel gain $a = 0$, then we can achieve $3$ degrees of freedom.
\end{rem}

\section{Comparing Lattice Coding and Han Kobayashi Coding}
In this section, we compare our layer lattice coding approach with the extension of Han-Kobayashi scheme for the three user symmetric Gaussian IC. In the Han-Kobayashi coding scheme for the two user IC \cite{HanKobayashi1981}, each transmitter splits its message into two parts, a private part and a common part. For decoding, each receiver decodes its message and the common message transmitted by the interfering transmitter. In the extension of this scheme to the three user IC, each transmitter splits its message into four parts - one private part and three common parts. For instance, transmitter $1$ splits its message $m_1$ into four parts - 1) $m_{11}$, the private part, 2) $m_{12}$, the common part which is also decoded by receiver $2$, 3) $m_{13}$, the common part which is also decoded by receiver $3$ and 4) $m_{123}$, the common part which is decoded by receivers $2$ and $3$. We restrict ourselves to Gaussian codebooks when considering the Han-Kobayashi coding strategy. Finally, as the channel is symmetric, we compare the maximum symmetric rate that can be achieved using the two approaches. In the next Lemma, we derive a symmetric rate point that can be achieved using the layered lattice coding approach for the three user symmetric Gaussian IC with cross channel gain $a$ and power constraint $P$. We define $P_a^N$ as follows
\begin{equation}
P_a^N \triangleq \left\{\begin{array}{l}(a^2-1)\frac{(2a^4-a^2)^N-1}{2a^4-a^2-1}, \textrm{ if } a^2 \geq 2 \vspace{0.1cm}\\
\frac{1-a^2}{2a^4}\frac{\left(\frac{1+a^2}{2a^4}\right)^{N_2}-1}{\left(\frac{1+a^2}{2a^4}\right)-1}, \textrm{ if } a^2 \leq \frac{1}{3}\end{array}\right..
\end{equation}
Let $\mathcal{R}_{HK}(P, \sigma^2, a)$ denote the maximum symmetric rate that can be obtained by Han-Kobayashi coding scheme in a three user symmetric Gaussian IC with power constraint $P$, noise at the receiver $\sigma^2$ and the cross channel gain equal to $a$.

\begin{figure}[hbtp]
\centering
\includegraphics[width=3.5 in]{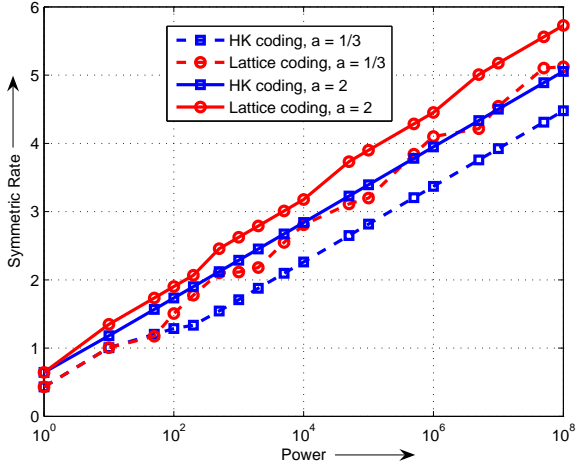}
\caption{Comparing Han-Kobayashi and Layered Lattice Coding for $a = 2.5$}
\label{fig : HK vs lattice 1}
\end{figure}

\begin{lem}\label{lem : symmetric rate}
Consider a symmetric three user Gaussian IC with cross channel gain $a$ and power constraint $P$ at each transmitter. Then\\
a) if $a^2 \geq 2$ and $P \leq a^2 - 1$, each user can achieve a symmetric rate given by
\begin{equation}
R_{sym} = \frac{1}{2}\log(P).
\end{equation}
b) If $a^2 \geq 2$, and there exists integer $N_1 > 0$ such that
\begin{displaymath}
P_a^{N_1} < P < P_a^{N_1+1},
\end{displaymath}
then each user can achieve a symmetric rate given by
\begin{equation}
R_{sym} = \frac{N_1}{2}\log(a^2-1) + \frac{1}{2}\log\left(1 + \frac{(2a^2+1)(P - P_a^{N_1})}{1 + (2a^2+1)P_a^{N_1}}\right).
\end{equation}
c) If $a^2 \geq 2$, and there exists integer $N_1 > 0$ such that $P_a^{N_1} = P$, then each user can achieve a symmetric rate given by
\begin{equation}
R_{sym} = \frac{N_1}{2}\log(a^2-1).
\end{equation}
d) If $a^2 \leq \frac{1}{3}$ and $P \leq \frac{1-a^2}{2a^4}$, each user can achieve a symmetric rate of
\begin{equation}
R_{sym} = \mathcal{R}_{HK}(P, a, 1).
\end{equation}
e) If $a^2 \leq \frac{1}{3}$ and there exists integer $N_2 > 0$ such that
\begin{displaymath}
P_a^{N_2} < P < P_a^{N_2+1}
\end{displaymath}
then each user can achieve a symmetric rate of
\begin{eqnarray}
\begin{array}{l}
R_{sym} = \max_{i = N_2 - 1 : N_2} \frac{i}{2}\log\left(\frac{1-a^2}{2a^2}\right) + \\
 \qquad\qquad\qquad\quad\mathcal{R}_{hk} (P - P_a^i, (2a^2+1)P_a^i, a)\end{array}
%R_{sym} &=& \frac{N_2-1}{2}\log\left(\frac{1-a^2}{2a^2}\right) + \vspace{0.15cm}\\
%&&
%\frac{1}{2}\log\left(\frac{P - P_a^{N_2-1}}{1 + P_a^{N_2-1} + 2a^2 P}\right).\end{array}
\end{eqnarray}
f) If $a^2 \leq \frac{1}{3}$ and there exists integer $N_2 > 0$ such that $P = P_a^{N_2}$, then each user can achieve a symmetric rate given by
\begin{equation}
R_{sym} = \frac{N_2}{2}\log\left(\frac{1-a^2}{2a^2}\right).
\end{equation}
\end{lem}
The proof of the above Lemma is very similar to the proof of Theorem 1 and is omitted here. It should be noted that for cases (b) and (e) in the lemma, we use Han-Kobayashi style encoding and decoding for the first layer of the codebook. This is because, the power allocated to this level is not sufficient enough to reap the advantages of lattice coding.  Figure \ref{fig : HK vs lattice 1} compares the symmetric rate point achievable using the layered lattice coding approach with the maximum symmetric rate that can be achieved using Han-Kobayashi scheme for $a = 2.5$ and $a = \frac{1}{3}$. Note that in our layered lattice coding approach, we restrict ourselves to identical power splitting approach by all the transmitters. This can be generalized to different power splitting schemes and can possibly give better results. However, it is interesting to note that even a possibly suboptimal lattice coding scheme significantly outperforms the extension of the Han-Kobayashi scheme.

This shows that while the Han-Kobayashi coding scheme (message splitting and random coding) is optimal to within one bit for a two user Gaussian IC \cite{EtkinTseWang2007}, it is not optimal even in terms of degrees of freedom for larger ICs with more than two transmitter-receiver pairs. The results of the paper in fact suggest that the optimal strategy might in fact be interference alignment through lattice coding. Lattice coding while allowing the interference to be decoded without decoding the interfering messages places fewer constraints on the rates of the interfering users. In particular, it eliminates the MAC type constraints that arise when decoding the interfering messages separately.

\section{Generalizations to non-symmetric channels}
In this section, we generalize the results of Section IV to a class of non-symmetric three user Gaussian ICs. We consider channel whose channel matrix $H \in \mathcal{H}_1$.
As $H \in \mathcal{H}_1$, there exists integers $p$ and $q$ such that $GCD(p, q) = 1$ and
\begin{equation}\label{eqn : channel matrix condition 2}
\frac{h_{12}}{h_{21}}\times \frac{h_{23}}{h_{32}}\times \frac{h_{31}}{h_{13}} = \frac{p}{q}.
\end{equation}
In the next Lemma, we provide a generalization of the ``very strong" interference result in \cite{Sridharan-Globecom} to non-symmetric channels.

\begin{lem}
Consider a three user Gaussian IC, whose channel matrix $H \in \mathcal{H}_1$ and whose channel gains satisfy (\ref{eqn : channel matrix condition 2}). The power constraints at the transmitters are $P_1, P_2, P_3$ and the noise variances at the receivers are $\sigma_1^2, \sigma_2^2$ and $\sigma_3^2$. If the channel gains satisfy one of the following three equations
\begin{eqnarray}\label{eqn : strong interference condition 1}
h_{12}^2 \geq p^2 \frac{P_1 + \sigma_1^2}{\sigma_2^2}, \quad h_{13}^2 \geq \frac{P_1 + \sigma_1^2}{\sigma_3^2}\nonumber\\
h_{21}^2 \geq q^2 \frac{P_2 + \sigma_2^2}{\sigma_1^2}, \quad h_{23}^2 \geq \frac{P_2 + \sigma_2^2}{\sigma_3^2}\\
h_{31}^2 \geq \frac{P_3 + \sigma_3^2}{\sigma_1^2}, \quad h_{32}^2 \geq \frac{P_3 + \sigma_3^2}{\sigma_2^2}\nonumber
\end{eqnarray}
(or)
\begin{eqnarray}\label{eqn : strong interference condition }
h_{12}^2 \geq \frac{P_1 + \sigma_1^2}{\sigma_2^2}, \quad h_{13}^2 \geq \frac{P_1 + \sigma_1^2}{\sigma_3^2}\nonumber\\
h_{21}^2 \geq \frac{P_2 + \sigma_2^2}{\sigma_1^2}, \quad h_{23}^2 \geq p^2 \frac{P_2 + \sigma_2^2}{\sigma_3^2}\\
h_{31}^2 \geq \frac{P_3 + \sigma_3^2}{\sigma_1^2}, \quad h_{32}^2 \geq q^2 \frac{P_3 + \sigma_3^2}{\sigma_2^2}\nonumber
\end{eqnarray}
(or)
\begin{eqnarray}\label{eqn : strong interference condition 3}
h_{12}^2 \geq \frac{P_1 + \sigma_1^2}{\sigma_2^2}, \quad h_{13}^2 \geq q^2 \frac{P_1 + \sigma_1^2}{\sigma_3^2}\nonumber\\
h_{21}^2 \geq \frac{P_2 + \sigma_2^2}{\sigma_1^2}, \quad h_{23}^2 \geq \frac{P_2 + \sigma_2^2}{\sigma_3^2}\\
h_{31}^2 \geq p^2 \frac{P_3 + \sigma_3^2}{\sigma_1^2}, \quad h_{32}^2 \geq \frac{P_3 + \sigma_3^2}{\sigma_2^2}\nonumber
\end{eqnarray}
then, the users can achieve rates given by
\begin{equation}
R_i \leq \frac{1}{2}\log\left(\frac{P_i}{\sigma_i^2}\right), \quad i\in\{1,2,3\}.
\end{equation}
That is, each user can achieve rates within half a bit of their maximum possible rates.
\end{lem}
The proof of the lemma is very similar to the proof of \cite[Theorem 2]{Sridharan-Globecom}. We describe below a brief outline of the proof.when the channel gains satisfy (\ref{eqn : strong interference condition 1}).

\textit{Outline of the Proof: } The encoding and decoding scheme are similar to those used in \cite[Theorem 2]{Sridharan-Globecom}. Each transmitter uses lattice codes to transmit their messages. For $i \in \{1,2,3\}$, transmitter $i$ chooses lattice $\Lambda_i$ generated using construction A described in (\ref{eqn : construction A}) or in \cite{Loeliger}. To ensure that the interference lattices are aligned at each receiver, we choose the lattices so that
\begin{equation}
h_{12}\Lambda_2 = ph_{13}\Lambda_3,\  h_{21}\Lambda_1 = qh_{23}\Lambda_3, \ h_{31}\Lambda_1 = h_{32}\Lambda_2.
\end{equation}
Let $V_i$ denote the volume of the Voronoi region of lattice $\Lambda_i$. The shaping region chosen by user $i$ is a $n$ dimensional sphere of radius $\sqrt{nP_i}$ denoted by $S_i$. We denote the volume of $S_i$ by $V_{S_i}$. Then, we choose the following relationship between $V_i, V_{S_i}$ and $R_i$
\begin{equation}
V_i = \frac{V_{S_i}}{2^{nR_i}}, \quad i \in \{1,2,3\}.
\end{equation}
The shifts for transmitter $i$ denoted by $t_i$ is chosen such that the codebook $\mathcal{C}_i$ given by $\mathcal{C}_i = (\Lambda_i + t_i) \cap S_i$ has at least $2^{nR_i}$ codewords. For decoding, receiver $i$ first decodes the total interference it sees and then decodes its own message. We describe the rate constraints involved in successful decoding at receiver $1$.

The total interference seen by receiver $1$ is given by
\begin{equation}
I_1 = h_{12} X_2 + h_{13} X_3
\end{equation}
Hence, the interference signal is an element of lattice $\Lambda_{i1} = h_{13}\Lambda_3$. In decoding the total interference, receiver $1$ sees a noise of $X_1 + Z_1$ whose power $\leq P_1 + \sigma_1^2$. Therefore, the interference can be successfully decoded if
\begin{eqnarray}
R_2 \leq \frac{1}{2}\log\left(\frac{h_{12}^2P_2}{p^2 (P_1 + \sigma_1^2)}\right)\\
R_3 \leq \frac{1}{2}\log\left(\frac{h_{13}^2P_3}{P_1 + \sigma_1^2}\right).
\end{eqnarray}
After decoding the interference, receiver $1$ decodes its message. The probability of decoding error decays with $n$ if
\begin{equation}
R_1 \leq \frac{1}{2}\log\left(\frac{P_1}{\sigma_1^2}\right).
\end{equation}
The analysis for receivers $2$ and $3$ are similar to that of receiver $1$ and the details are omitted here. If the channel gains satisfy (\ref{eqn : strong interference condition 1}), then the rate constraints due to decoding interference are less binding than the rate constraints due to decoding the message at each receiver. Hence, the users can achieve rates given by
\begin{displaymath}
R_i \leq \frac{1}{2}\log\left(\frac{P_i}{\sigma_i^2}\right), \quad i \in \{1,2,3\}.
\end{displaymath}
This completes the outline of the proof when the channel gains satisfy (\ref{eqn : strong interference condition 1}). Next, we analyze the degrees of freedom of non-symmetric ICs.

\subsection{Degree of Freedom Analysis}
In this section, we briefly analyze the degrees of freedom of three user non-symmetric Gaussian ICs. We use the same layered lattice coding scheme that we used for the symmetric case. 
%However, we are not able to derive closed form expressions for the total degrees of freedom achievable.
To present the main ideas, we consider an example three user Gaussian IC with channel matrix given by
\begin{equation}\label{eqn : toy example channel}
H = \left(\begin{array}{ccc}1 & a_1 & a_1\\ a_2 & 1 & a_2 \\ a_3 & a_3 & 1\end{array}\right),
\end{equation}
where $a_1^2, a_2^2, a_3^2 \geq 2$. Without loss of generality we assume $a_1 \leq a_2 \leq a_3$. The analysis for other channel matrices in $H \in \mathcal{H}_1$ are similar to the one presented and is omitted here. We describe the encoding and decoding strategy below:

For $j \in \{1,2,3\}$, transmitter $j$ communicates message $m_j \in \{1, \ldots, 2^{nR_j}\}$ to receiver $j$. Transmitter $j$ splits its message into $N$ parts - $m_{j_1}, m_{j_2}, \ldots, m_{jN}$ such that rate $R_{ji}$ is associated with the $i^{th}$ sub-message. For $i \in \{1, \ldots, N\}$, transmitter $j$ encodes message $m_{ji}$ into codeword $X_{ji}^n$ and transmits $X_j^n = \sum_{i=1}^N X_{ji}^N$. Also, transmitter $j$ assigns power $P_{ji}$ to encode its $i^{th}$ sub-message.

Encoding Strategy: Each transmitter encodes all its sub-messages using lattice coding, and chooses lattices $\Lambda_1, \ldots, \Lambda_N$. Transmitter $j$ chooses shifts $s_{j1}, \ldots, s_{jN}$ and spherical shaping regions $S_{j1}, \ldots, S_{jN}$. The codebook for the $i^{th}$ sub-message at transmitter $j$ is denoted by $\mathcal{C}_{ji}$ and is given by $\mathcal{C}_{ji} = (\Lambda_i + s_{ji}) \cap S_{ji}$.

\begin{figure}[hbtp]
\centering
\includegraphics[width=3.5 in]{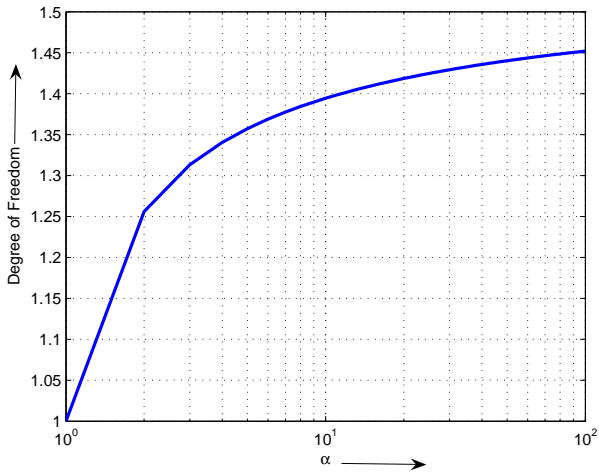}
\caption{Degree of Freedom for an example channel: $a_1 = 2\alpha, a_2 = 3\alpha, a_3 = 4\alpha$}
\label{fig : degree of freedom non symmetric}
\end{figure}

Decoding Strategy: The received signal at receiver $j$ is given by
\begin{displaymath}
Y_j^n = \sum_{i=1}^N X_{ji}^n + \sum_{l=1, l\neq j}^3 \sum_{i=1}^N a_j X_{li}^n + Z_j^n.
\end{displaymath}
We denote the interference at receiver $j$ due to the $i^{th}$ sub-message from the other transmitter by $I_{ji}^n$ and is given by
\begin{displaymath}
I_{ji}^n = \sum_{l=1,l\neq j}^3 a_j X_{li}^n.
\end{displaymath}
The decoding process at receiver $j$ proceeds through $N$ stages. At stage $i$, receiver $j$ first decodes interference $I_{ji}^n$ and then decodes its sub-message $m_{ji}$. In decoding interference $I_{ji}^n$, receiver $j$ sees an effective noise power of
\begin{equation}
\sigma_{ji}^2 = 1 + P_{ji} + \sum_{l = i+1}^N P_{jl} + \sum_{l=1, l\neq j}^{3}\sum_{k=i+1}^{N} a_j^2 P_{lk}.
\end{equation}
In decoding message $m_{ji}$, receiver $j$ sees an interference plus noise power of
\begin{equation}
\sigma_{m_{ji}}^2 = \sum_{l = i+1}^N P_{jl} + \sum_{l=1, l\neq j}^{3}\sum_{k=i}^{N} a_j^2 P_{lk}.
\end{equation}
The choice of lattices $\Lambda_i$, shifts $s_{ji}$ and shaping regions $S_{ji}$ are similar to those described in Theorem 1 and the details are omitted here. We choose the powers $P_{ji}$ such that the ``very strong" interference condition is satisfied at every decoding stage. That is, at stage $i$, the rate constraints on $R_{ji}$ due to decoding interference $I_{li}^n$ at receiver $l$ is less binding than the constraint imposed due to decoding message $m_{ji}$ at receiver $j$. Hence, we choose powers such that
\begin{eqnarray}
\begin{array}{c}
P_{1i} = \min(a_1^2 \sigma_{m_{2i}}^2 - \sigma_{m_{1i}}^2, a_1^2 \sigma_{m_{3i}}^2 - \sigma_{m_{1i}}^2)\\
P_{2i} = \min(a_2^2 \sigma_{m_{1i}}^2 - \sigma_{m_{2i}}^2, a_2^2 \sigma_{m_{3i}}^2 - \sigma_{m_{2i}}^2)\\
P_{3i} = \min(a_3^2 \sigma_{m_{1i}}^2 - \sigma_{m_{3i}}^2, a_3^2 \sigma_{m_{2i}}^2 - \sigma_{m_{3i}}^2).\end{array}
\end{eqnarray}
The rate achieved by user $j$ at stage $i$ is given by
\begin{equation}
R_{ji} = \frac{1}{2}\log\left(\frac{P_{ji}}{\sigma_{m_{ji}}^2}\right).
\end{equation}
The total power used by transmitter $j$ is given by
\begin{displaymath}
P_j = P_{j1} + P_{j2} + \ldots P_{jN}.
\end{displaymath}
The total degrees of freedom then satisfies
\begin{displaymath}
\mathcal{D}_{sum} \geq \limsup_{P_1 + P_2 + P_3 \rightarrow \infty}\frac{\sum_{j=1}^3\sum_{i=1}^N R_{ji}}{\frac{1}{2}\log(P_1+P_2+P_3)}.
\end{displaymath}
However, unlike the symmetric channel case in Theorem 1, we have not been able to derive closed form expressions for the total degrees of freedom achievable for non-symmetric channels. We illustrate the total degrees of freedom achieved for an example channel (derived numerically) in Figure \ref{fig : degree of freedom non symmetric}.
The degree of freedom analysis for other non symmetric three user Gaussian ICs with channel matrix $H \in \mathcal{H}_1$ follows along the same lines as the analysis for the channel given by (\ref{eqn : toy example channel}).

\section{Conclusions}
In this paper, we proposed a layered lattice coding scheme for the three user Gaussian IC. In Theorem $1$, we showed that the layered lattice coding approach can achieve more than degree of freedom for symmetric channels. We also derived an analytical expression for the achievable degrees of freedom in terms of the channel gain $a$. In Section V, we compared the Han-Kobayashi coding strategy with our layered lattice coding scheme for symmetric ICs. We showed that significant rate benefits can be achieved by decoding the interference rather than decoding the interfering messages.  Finally, in Section VI, we extended the ``very strong " interference result of \cite{Sridharan-Globecom} to a class of non-symmetric channels and used this to generalize the layered lattice coding approach to that class of channels.

\end{document}